\documentclass{vgtc}                          

\graphicspath{{figures/}{pictures/}{images/}{./}} 

\usepackage{times}                     

\usepackage{tabu}                      
\usepackage{booktabs}                  
\usepackage{lipsum}                    
\usepackage{mwe}                       
\usepackage{balance}
\usepackage{mathptmx}                  
\usepackage{amsmath}
\usepackage{paralist,bbding,pifont}
\usepackage{authblk}
\newcommand{\prefix}{\textcolor[RGB]{0, 102, 204}}

\newcommand{\ouyang}{\textcolor{black}}

\onlineid{0}

\vgtccategory{Research}

\vgtcinsertpkg

\title{A Two-Phase Visualization System for Continuous Human-AI Collaboration in Sequelae Analysis and Modeling}

\makeatletter

\renewcommand\AB@affilsepx{, \protect\Affilfont}
\makeatother
\author[1]{Yang Ouyang\thanks{e-mail: ouyy@shanghaitech.edu.cn}}
\author[2]{Chenyang Zhang\thanks{e-mail: zhang414@illinois.edu}}
\author[1]{He Wang\thanks{e-mail: wanghe1@shanghaitech.edu.cn}}
\author[3]{Tianle Ma\thanks{e-mail: ma.tianle@zshospital.sh.cn}}
\author[3]{Chang Jiang\thanks{e-mail: cjiang\_fdu@yeah.net}}
\author[3]{Yuheng Yan, Zuoqin Yan\thanks{e-mail: \{yan.yuheng,yan.zuoqin\}@zshospital.sh.cn}}
\author[4]{Xiaojuan Ma\thanks{e-mail: mxj@cse.ust.hk}}
\author[5]{Chuhan Shi\thanks{e-mail: chuhanshi@seu.edu.cn}}
\author[1]{Quan Li\thanks{e-mail: liquan@shanghaitech.edu.cn (Corresponding author)}}

\affil[1]{ShanghaiTech University}
\affil[2]{University of Illinois at Urbana-Champaign}
\affil[3]{Zhongshan Hospital Fudan University}
\affil[4]{Hong Kong University of Science and Technology}
\affil[5]{Southeast University}

\teaser{

    \centering
     \vspace{-8mm}
      \includegraphics[width=\linewidth]{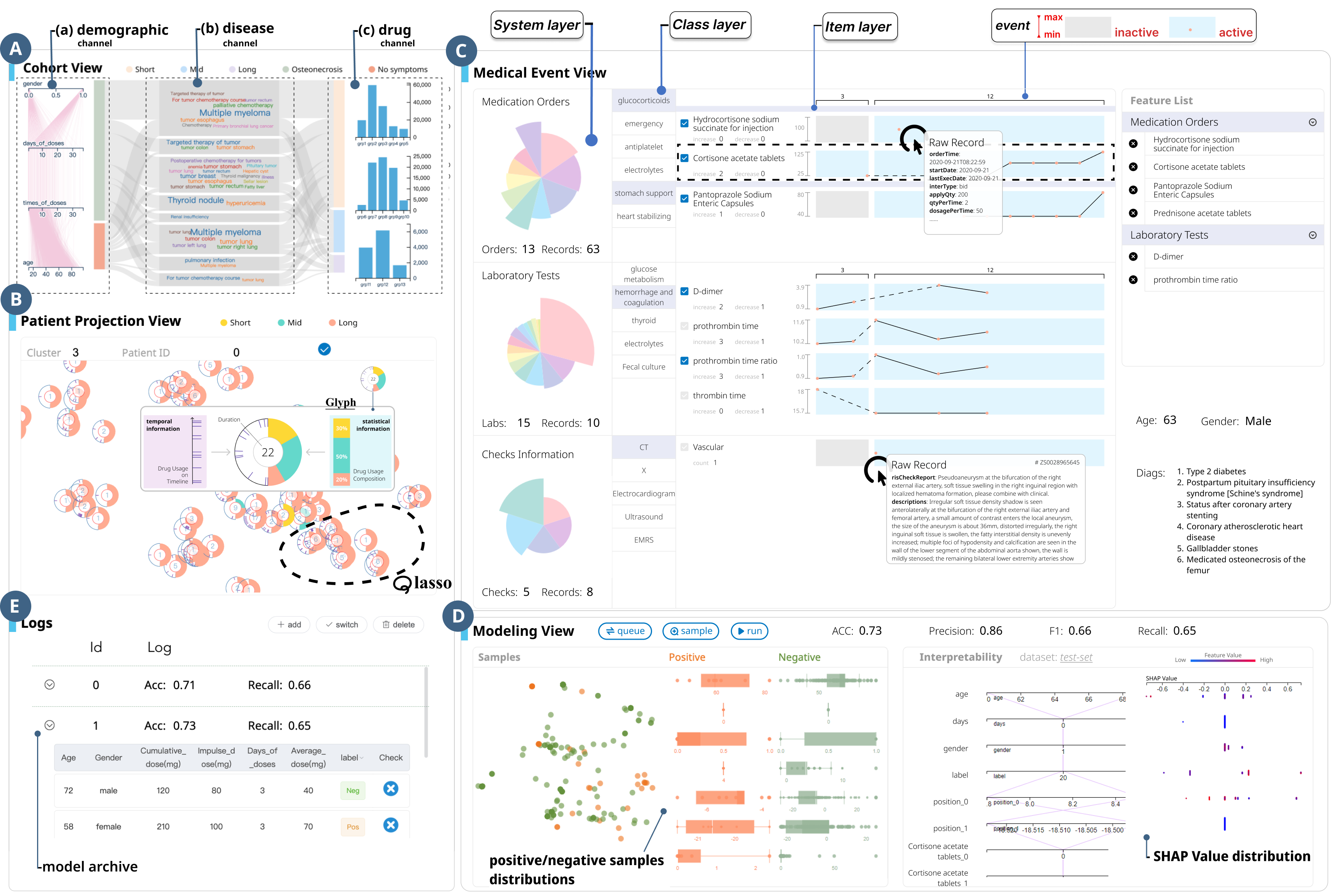}
      \vspace{-6mm}
      \caption{Our system overview: \textbf{Phase I} includes (A) \textit{Cohort View} for understanding drug event and disease progression relationships, (B) \textit{Patient Projection View} to explore specific patient cohort characteristics, and (C) \textit{Medical Event View} for detailed visualization of patient medical events. \textbf{Phase II} comprises (D) \textit{Modeling View} for iterative AI model development and performance evaluation, and (E) \textit{Logs View} for maintaining iteration records of models and associated data.}
      \label{fig:teaser}
  
}
\abstract{
    In healthcare, AI techniques are widely used for tasks like risk assessment and anomaly detection. Despite AI's potential as a valuable assistant, its role in complex medical data analysis often oversimplifies human-AI collaboration dynamics. To address this, we collaborated with a local hospital, engaging six physicians and one data scientist in a formative study. From this collaboration, we propose a framework integrating two-phase interactive visualization systems: one for Human-Led, AI-Assisted Retrospective Analysis and another for AI-Mediated, Human-Reviewed Iterative Modeling. This framework aims to enhance understanding and discussion around effective human-AI collaboration in healthcare.}    


\begin{document}
\firstsection{Introduction}

\maketitle
\par In healthcare, AI integration offers substantial advancements, particularly in tasks like risk assessment and anomaly detection~\cite{tigga2020prediction,verma2023rethinking}. AI is acknowledged for its ability to uncover nuanced details and initiate collaborative efforts. However, existing studies often view fully developed AI models merely as assistants or reminders~\cite{tigga2020prediction,Nagar2021MEDICAREAA}, focusing narrowly on task-specific performance within specialized domains. This approach overlook AI's limitations in complex medical data scenarios. Previous research has predominantly concentrated on protopathic diseases, characterized by distinct, short-term symptoms~\cite{tigga2020prediction,2017Visual}. However, in the context of complex medical data scenarios, such as those involving multifaceted diseases, the intricate nature of the data demands more sophisticated processing and analysis capabilities. Enhancing AI's effectiveness in such scenarios requires precise feature definition and meticulous input labeling by medical experts. Therefore, understanding the roles of humans and AI, including role transfer and task allocation across different analysis stages, is critical for optimizing collaborative outcomes.
\par In response, we partnered closely with a local hospital, embedding ourselves in the daily routines of domain experts. Initially focusing on hormone-related medical records, we conducted a formative study involving six physicians and one data scientist. Through insightful interviews, we explored their needs and expectations for sequelae analysis, emphasizing roles and tasks across different analysis stages. This collaborative process identified seven design prerequisites structured into two levels: \textbf{\textit{Human-Led, AI-Assisted Retrospective Analysis}} and \textbf{\textit{AI-Mediated, Human-Reviewed Iterative Modeling}}. Subsequently, we devised a framework with two phases aligned with these prerequisites. \textbf{Phase I} introduced a retrospective visualization system for interactive data analysis, fostering co-design of input features and preparing for Human-AI collaboration. \textbf{Phase II} extended these insights, integrating AI models into the final analytical process based on medical experts' feedback.


\section{Related Work}
\subsection{Human-AI Collaboration}
\par In the realm of Human-AI collaboration, three prevalent forms have been identified~\cite{cai2019human}. The first involves \textit{AI-assisted decision-making}, where AI provides decision suggestions while humans make the final decisions~\cite{bansal2021does,cai2019hello,zhang2020effect}. For example, Cai et al.~\cite{cai2019hello} explored medical practitioners' onboarding needs with diagnostic AI assistance, while Zhang et al.~\cite{zhang2020effect} studied the impact of displaying AI prediction confidence on human trust in AI. The second form is \textit{Human-in-the-loop}, where human input enhances AI performance. Lee et al.~\cite{lee2021human} designed a collaborative approach allowing therapists to review AI outputs and offer feedback for improvement. Interactive ML also falls into this category, involving human participation in AI model predictions to enhance performance~\cite{lai2022human,ma2022modeling,mishra2021designing}. The third form is \textit{joint action}, wherein humans and AI collaborate toward a shared goal~\cite{ashktorab2020human,zuckerman2022tangible}. Research in this domain primarily explores appropriate task allocation between humans and AI. For instance, Lai et al.~\cite{lai2020chicago,lai2019human} proposed allocation schemes for deceptive review detection, distributing work between humans and AI based on varying degrees of human involvement.
\par Our work integrates human domain expertise with AI's capabilities. In \textbf{Phase I}, we focus on retrospective analysis, establishing links between diseases, medical incidents, and outcomes, encompassing \textit{AI-assisted} risk identification. In \textbf{Phase II}, we empower users to participate in a \textit{Human-in-the-loop} process, aiding them in identifying suitable sample cohorts for AI modeling.

\subsection{Medical Data Visualization}
\par Electronic Health Records (EHRs) often contain event sequence data. Traditional visualization approaches for such data include timelines, point or interval plots~\cite{plaisant2003lifelines,shahar2006distributed}, often incorporating innovative designs like glyphs and tables~\cite{cao2011dicon,ghassemi2018clinicalvis}. Some studies represent event sequences using tree-based structures~\cite{liu2017coreflow} or Sankey-based methods~\cite{gotz2014decisionflow}. Recent work, like ThreadState~\cite{wang2021threadstates}, introduced novel matrix and scatter plot combinations to reveal disease progression patterns. When dealing with multiple patient records or different cohorts, glyphs are commonly used to represent records with various variables and features~\cite{cheng2021vbridge,2015TimeSpan}. For instance, some studies combine and integrate multiple patient records for comparing symptom evolution~\cite{2008Aligning}, while others develop tools like \textit{TimeSpan}~\cite{2015TimeSpan} to facilitate the exploration of multidimensional data on specific patient groups, enhancing decision-making and patient care. Unlike efforts primarily focused on protopathic diseases, our study targets various EHRs related to sequelae. We aim to assist physicians in comprehending and assessing sequelae-related medical data by starting with a defined patient cohort and guiding them through individual patient data, using detailed visual designs.

\section{Formative Study}

\label{subsection:designs}
\par During our three-month formative study, we collaborated closely with seven experts (\textbf{E1-E7}) from a renowned local hospital. \textbf{E1-E6} possess extensive clinical and research expertise in orthopedic surgery, spinal cord injuries, and degenerative bone and joint diseases. \textbf{E7}, from the Big Data department, contributed to designing patient groups and analyzing data for real-world research on subordinate cohorts, focusing on hormone-related osteonecrosis. Our study prioritized monitoring glucocorticoid-centered medications and early identification of osteonecrosis risk. \ouyang{Unlike diseases with distinct, short-term symptoms, hormone-related diseases often exhibit subtle characteristics, ambiguous concepts and intricate pathogenesis.} We conducted semi-structured interviews, lasting about 30 minutes each, with individual experts to explore their experiences in assessing sequelae risks and timelines, including planning strategies and processes. Additionally, detailed discussions with experts were conducted to explore the concepts of roles and tasks across various analysis stages.
\par \ouyang{We first identified key challenges in traditional practices, including data quality issues, ambiguity in identifying risk factors, and limitations in patient sample selection. These problems complicate effective analysis, hinder precise risk factor determination, and limit generalizability. Furthermore, AI models often lack transparency and interpretability, making it difficult for medical professionals to trust their recommendations. Additionally, collaboration with AI is hindered by a lack of familiarity with developing and validating these models.} We then presented the ultimate design requirements, covering the original specifications and addressing challenges from interviews and iterative design insights.
We organized these requirements into two levels: \textbf{\textit{Human-Led, AI-Assisted Retrospective Analysis}} (\textbf{Level I}) and \textbf{\textit{AI-Built, Human-Reviewed Iterative Modeling}} (\textbf{Level II}), each delineating unique roles and responsibilities for the physicians.


\par \prefix{\texttt{\textbf{[Level I]}}} \textbf{DR1. Offer a concise summary of hormone-related EHR data.} Collaborating physicians emphasized the importance of quickly grasping key aspects of the data, as there could be potential data noise related to sequelae in medical records. In particular, one physician pointed out that earlier data might lack structure and specificity, potentially limiting its research utility. Therefore, a straightforward and clear summary is considered essential to provide physicians with a comprehensive view of the hormone-related medical data.

\par \prefix{\texttt{\textbf{[Level I]}}} \textbf{DR2. Identify target patient population.}  Following the conventional approach of domain experts, it is advantageous to focus on particular medical indicators and perform targeted analyses on the relevant population. E1 recommended considering demographic factors like age, gender, medication type, and patients' protopathy or sequelae as valuable contributors to this effort. Furthermore, E2 expressed a strong interest in investigating the potential connection between glucocorticoid medications and osteonecrosis. This emphasis was particularly directed towards patients using glucocorticoids and those afflicted with osteonecrosis.

\par \prefix{\texttt{\textbf{[Level I]}}} \textbf{DR3. In-depth examination of individual patient medical events.} Our discussions with physicians highlighted their significant concern regarding information bias stemming from subjective interpretation. As a result, it has become imperative to develop a detailed visualization that precisely depicts the entire trajectory of an individual patient's medical events. E2 underscored the importance of a comprehensive medication record display to facilitate a meticulous examination of the original data records. 


\par \prefix{\texttt{\textbf{[Level I]}}} \textbf{DR4. Identify risk factors.} The retrospective analysis of medical data related to sequelae serves to identify features that can differentiate sequelae and facilitate the construction of ML models. As E1 expounded, ``\textit{because patients' exposure to other risk factors in non-hospital settings is unknown, retrospective analysis of real-world medical data is fundamental and crucial to developing such models.}'' At this stage, the approach should center on cross-sectional and longitudinal comparisons of individual patients' medical events and conditions to reveal potential differences between various patient cohorts. 

\par \prefix{\texttt{\textbf{[Level II]}}} \textbf{DR5. Mandate the automated extraction and display of positive and negative samples.} 
To address potential analysis imbalance, physicians utilized Propensity Score Matching (PSM)~\cite{austin2011introduction}. However, it's crucial to note, as emphasized by E5, that PSM may not effectively tackle issues related to selection bias or omitted variables and might exclude samples lacking specific characteristics. 
Physicians need an automated solution that simplifes the identification of appropriate patient groups while ensuring a balanced distribution of confounders (or covariates) between positive and negative groups. Also, physicians require a clear understanding of characteristic distribution within these samples. E2 elaborated, stating, ``\textit{we must visualize the range and variation of each characteristic within both the positive and negative samples, and discern the differences between them. This can help us determine which features hold greater importance or influence over the outcomes.}''

\par \prefix{\texttt{\textbf{[Level II]}}} \textbf{DR6. Ensure transparency and interpretability in AI models.} It is imperative that the AI model possesses the qualities of transparency and interpretability. This implies that the model should be capable of providing comprehensible and meaningful rationales for its predictions, including aspects such as the weights or importance of each feature, decision rules or logic, as well as the level of uncertainty or confidence. As stated by E4, ``\textit{Understanding the inner workings of the model and the reasoning behind its predictions is a matter of great interest to us. It not only aids in model validation but also enhances our understanding of the risk factors associated with hormone-related osteonecrosis.}''

\par \prefix{\texttt{\textbf{[Level II]}}} \textbf{DR7. Actively participate, receive feedback, and engage in iterative interaction.} In the course of analyzing hormone-related osteonecrosis, experts must meticulously track every aspect of the process, including sample and feature selection, model training and evaluation, and the subsequent interpretation and validation of results. It is essential for the system to meticulously record and track all these procedures and their associated data, enabling experts to effortlessly retrieve and review them whenever necessary. E1 succinctly conveyed the importance, stating, ``\textit{Despite our non-expertise in AI, we need a system that provides a clear understanding of how AI reaches its conclusions}''. 

\section{Phase I: Human-Led, AI-Assisted Retrospective Analysis}
\par
\begin{figure}[h]
  \centering
  \vspace{-4mm}
  \includegraphics[width=\linewidth]{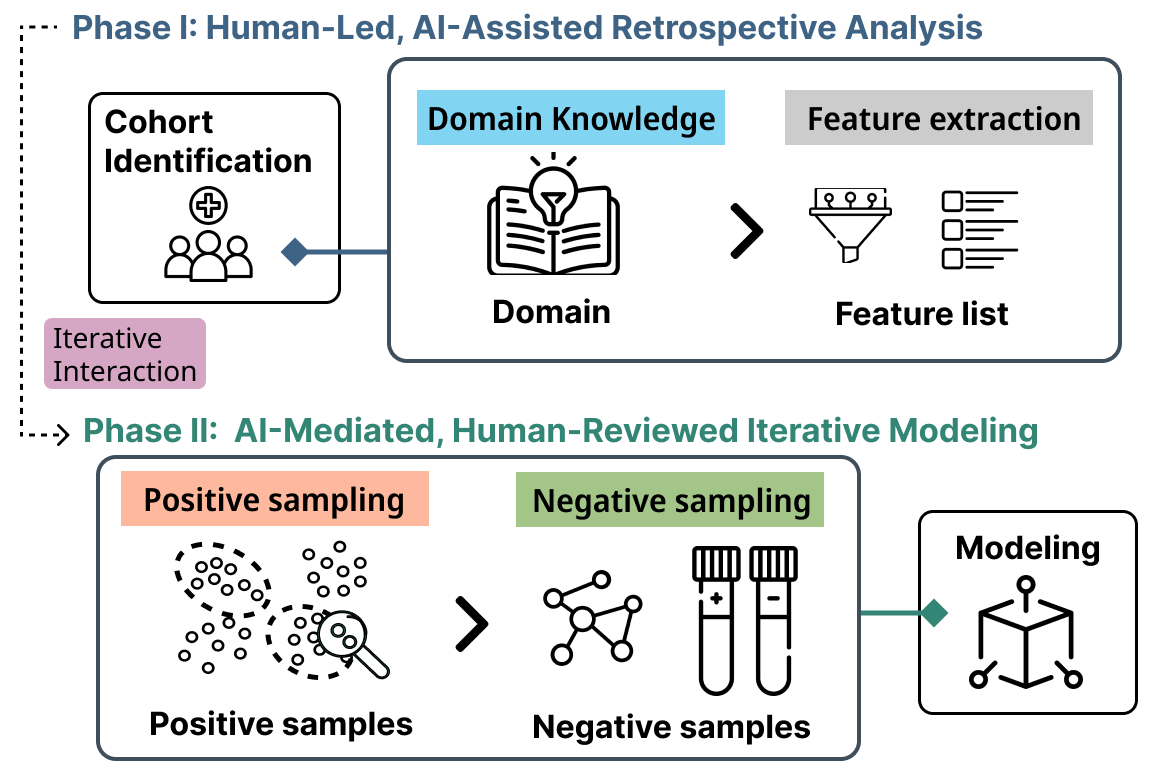}
  \vspace{-6mm}
    \caption{The two-phase framework overview.}
      \vspace{-6mm}
  \label{fig:framework_overview}
\end{figure}

 During \cref{fig:framework_overview} (\textbf{Phase I}), we utilized medical records associated with sequelae, enabling domain experts to streamline their analysis of drug and hormone side effects and identify significant risk factors affecting the diagnosis process (\textbf{DR1 - DR4}). In this phase, humans lead the viewing, screening, and retrospective analysis process, with AI aiding to enhance ease and efficiency.

\subsection{Data Processing and Characteristics}
\par The data was collected from the Health Information System (HIS) provided by the experts. \ouyang{We ensured patient confidentiality through rigorous data anonymization and obtained IRB approval from the Research Ethics Committee.} In addition to capturing essential details such as Personal Information and Admit and Discharge Time, the remaining data is categorized into five groups: (\textbf{DR1}): Primary Diagnosis\prefix{\texttt{\textbf{[G1]}}}, Laboratory Tests \prefix{\texttt{\textbf{[G2]}}}, Examination Information \prefix{\texttt{\textbf{[G3]}}}, Medication Orders \prefix{\texttt{\textbf{[G4]}}}, and Medical Records \prefix{\texttt{\textbf{[G5]}}}.

\subsection{Visualization for Phase I: Cohort - Projection - Event}
\par The core principle guiding \textbf{Phase I} design is to replicate the routine practices of medical professionals, facilitating retrospective analysis through three key features: \textit{Cohort} - \textit{Projection} - \textit{Medical Event}.

\textbf{Cohort View.}
To gain a more holistic understanding of the interactions between medications and diseases, we introduce the \textit{Cohorts View} (\cref{fig:teaser}(A)), which elucidates the connection between drug events and disease progression, encompassing primary disease and sequelae. This view comprises three key elements: the demographics channel (\cref{fig:teaser}(A-a)) on the left, the disease channel (\cref{fig:teaser}(A-b)) in the center, and the drug channel (\cref{fig:teaser}(A-c)) on the right. We employ a Sankey design to establish connections between these channels. In the demographics channel, we use parallel coordinates to present demographic data for two distinct populations: those with and without osteonecrosis. Moving to the middle channel, primary diseases are represented as clusters in word clouds, where the size of each cloud reflects the patient count. In the drug channel, we illustrate the three hormone classes, namely \textit{short-acting}, \textit{medium-acting}, and \textit{long-acting} based on their duration of action. Bar charts are employed to display the medication distribution within these classes. This design offers physicians a comprehensive overview of patient cohorts, their primary diseases, and available drug treatment options. Additionally, the \textit{Cohort View} enables users to select specific clusters of interest by clicking on corresponding rectangles, facilitating further exploration in alignment with \textbf{DR2}.

\par \textbf{Patient Projection View.} In collaboration with physicians and E7, we have gained insights into the common use of dimensionality reduction techniques in their daily data analysis routines. Hence, we have incorporated this approach into \textbf{Phase I} to support physicians in exploring the characteristics of specific patient cohorts. Once a particular type of patient cohort is identified within the \textit{Cohort View}, users can delve deeper into their characteristics using the \textit{Patient Projection View} (\cref{fig:teaser}(B)) (\textbf{DR2}). The foundational design principle behind the \textit{Patient Projection View} involves the application of classical dimensionality reduction techniques, including \textit{t-SNE}~\cite{van2008visualizing}, \textit{PCA}~\cite{shlens2014tutorial}, \textit{MDS}~\cite{carroll1998multidimensional}, and \textit{UMAP}~\cite{mcinnes2018umap}. These techniques are employed to create low-dimensional projections that retain local similarity, effectively representing neighborhood structures. In the construction of the \textit{Patient Projection View}, we utilize the patient's feature vector, denoted as $vec(P)$, which incorporates the aforementioned features. This approach enables a more comprehensive exploration of patient cohort characteristics.

\par In the \textit{Patient Projection View}, a novel glyph (\cref{fig:teaser}(B-Glyph)) is utilized to depict each patient. This glyph comprises a central text indicating the patient's duration. On the left semicircle, arcs symbolize the complete inpatient cycle for the respective patient, with small vertical lines denoting specific medication events during that period. The right semicircle illustrates the patient's medication history, color-coded to distinguish between hormone classes (short-acting, intermediate-acting, and long-acting). The angle of each sector represents the dose of the medication class. 

\textbf{Medical Event View.}
The \textit{Medical Event View} provides a comprehensive record of a patient's health status, empowering users to gain insights into disease progression through retrospective analysis and identify pertinent details (\textbf{DR3 -- DR4}). To facilitate the clear presentation of this information, we developed an interactive table-based dashboard that allows users to monitor extensive dataset data efficiently. As illustrated in~\cref{fig:teaser}(C), the table is divided vertically into three cells: \textit{Medication Order}, \textit{Laboratory Tests}, and \textit{Checks Information}. These cells share a common timeline in terms of position and scale, enabling users to make horizontal and vertical comparisons across the patient's health record. Within each cell, information is presented in three progressively detailed layers, proceeding from left to right.

\section{Phase II: AI-Mediated, Human-Reviewed Iterative Modeling}
\par In \textbf{Phase II}, we integrated AI models into the entire analysis workflow. The insights obtained from \textbf{Phase I} also play a crucial role in identifying critical features that could effectively distinguish hormone-related sequelae within the appropriate patient cohorts. These insights, in turn, guided the collaborative design of input features and sample datasets customized for the subsequent AI modeling stage (\textbf{DR5 - DR7}). In this phase, AI functions as the builder, constructing the model using high-quality data from \textbf{Phase I}, while humans assume the role of reviewers, evaluating the model's performance and interpretability, providing feedback, and iterating on the model as needed, as depicted in \cref{fig:framework_overview} (\textbf{Phase-II}).

\subsection{Backend Experiments}
\par \textbf{Identify Positive and Negative Samples.} In response to the shortcomings of PSM, we recognized the importance to thoroughly examine both the feature space and sample cohorts (\textbf{DR5}).
We adopted critical steps including feature transformation and positive sampling, followed by the adoption of negative sampling strategies. Through a series of experiments, we identified \textit{hard negative sampling}~\cite{xuan2020improved} as a superior sampling strategy compared to alternatives.

\par \textbf{Construct Interpretable AI Models.}
After thoroughly evaluating the performance and interpretability of multiple models, we determined that the \textit{RandomForest} model was the most suitable for our prediction tasks. Additionally, we leveraged Shapley values~\cite{NIPS2017_7062} at the instance level to estimate the importance of features in predicting sequela outcomes and to explore the relationship between the features selected by physicians in \textbf{Phase I} and the resulting predictions (\textbf{DR6}).

\subsection{Visualization for Phase II: Modeling - Log}
\par Building upon the interface of \textbf{Phase I} and incorporating the interactive approach outlined in \textbf{DR5-DR7}, we developed additional designs for \textbf{Phase II}, including two key features: \textit{Modeling} - \textit{Logs}.

\par \textbf{Modeling View.} In the previous \textit{Medical Event View}, if physicians find specific features sufficiently distinctive, they can include them in a feature list on the right-hand side, which is continuously updated and integrated into the \textit{Medical Event View} (as shown in \cref{fig:teaser}(C-Feature List)). Similarly, in the previous \textit{Patient Projection View}, if physicians wish to focus on particular patient groups, they can employ a lasso operation within the view (as depicted in \cref{fig:teaser}(B-lasso)). Subsequently, they can examine the projection of these selected samples and the distribution of features in the left-hand section of the \textit{Modeling View} (Samples, Positive, Negative) (as shown in \cref{fig:teaser}(D-left)). In this view, green and yellow dots indicate positive and negative patients, and physicians can choose points of interest to access specific patient features in the right box plots (addressing \textbf{DR5}). Physicians can conduct a comprehensive comparison of selected positive and negative patient cohorts across various feature dimensions within the right box plots. Each horizontal axis represents a feature dimension, and the box plot illustrates the distribution of feature values. Alternatively, physicians can add specific samples meeting their criteria to an ongoing sample list in the backend. The predictions and feature interpretability results are displayed on the right side of the interface, as shown in \cref{fig:teaser}(D-right), addressing both \textbf{DR6} and \textbf{DR7}. These visualizations include a parallel coordinates chart showing feature distributions within the test dataset and a beeswarm plot displaying SHAP values for each feature in the test samples.

\par \textbf{Logs View.}
\label{section:log_view}
We present the \textit{Logs View}, depicted in \cref{fig:teaser}(E), to maintain records of the models and their associated data for each iteration, fulfilling \textbf{DR7}. This view encapsulates two essential pieces of information in each row: 1) The training data utilized. 2) The model's performance in terms of prediction. For more in-depth insights into a specific round, users can click the expand button, which will reveal the pertinent details for that round and simultaneously update the \textit{Modeling View.}

\section{Expert Evaluation}
\par We conducted semi-structured interviews with experts to evaluate our approach for exploring hormone-related medical records and analyzing sequelae. Feedback was categorized into three main areas: \textit{perception of the two-phase co-design process}, \textit{adaptation of workflows} and \textit{concerns about collaborating with AI}. Experts generally viewed our design process favorably, highlighting support for drug and hormone side effects exploration, new perspectives on sequela analysis, and comprehensive trade-off evaluation. They appreciated the intuitive visual metaphors, iterative selection and update of features, and the integration of their expertise into the AI model. The collaborative approach stimulated creativity and provided valuable insights, while the novel system design facilitated a more thorough analysis compared to previous methods. Concerns included potential AI misguidance due to poor-quality data and excessive reliance on AI, but overall, experts found the process beneficial and expressed intentions to incorporate it into future research.

\section{Discussion, Limitation and Conclusion}
\par Our study introduces a two-phase framework that collaborates with medical experts to create a human-AI collaborative system. This system enhances interactive analysis and co-design of AI features and samples. Unlike previous studies focused on AI explainability~\cite{wysocki2023assessing}, our system improves analysis efficiency and quality through interactive visualization and active user support, addressing data quality and potential sample biases. Initially designed for sequelae data, the framework can extend to other domains like biology and chemistry, aiding drug discovery~\cite{kunduru2023machine} and protein analysis~\cite{li2022recent}. Data must be preprocessed into a specific JSON format with structural and functional features for drug or protein targets, focusing on tasks like predicting properties and toxicity~\cite{sadybekov2023computational}. The system's interaction design supports seamless exploration of cohort data and continuous AI collaboration in hypothesis testing. Modularizing core components allows customization for different domains. However, our work has limitations. EHRs may introduce biases, affecting accuracy and generalizability~\cite{rajkomar2019machine}. The focus on medication use and binary ``checks'' may limit comparisons, requiring further abstraction and visualization. Our evaluation relied on qualitative interviews; future implementations should involve a larger pool of external users for comprehensive testing.

\balance

\bibliographystyle{abbrv-doi}

\bibliography{template}
\end{document}